\documentstyle[12pt]{article}

 \newcommand{\be}{\begin{equation}}
 \newcommand{\ee}{\end{equation}}
 \newcommand{\ba}{\begin{eqnarray}}
 \newcommand{\ea}{\end{eqnarray}}
 
 \newcommand{\del}{\partial}

\newcommand{\lef}{\left}

\newcommand{\ri}{\right}

\newcommand{\cl}{{\cal L}}

\newcommand{\fr}{\frac}

\begin{document}

\begin{titlepage}

\topmargin -15mm

\rightline{\bf UFRJ-IF-FPC-010/96}

\vskip 10mm

\centerline{ \LARGE\bf Duality and an Operator Realization }
\vskip 2mm
\centerline{ \LARGE\bf for the Fermi-Bose Transmutation in 3+1 Dimensions}

    \vskip 2.0cm

    \centerline{\sc E.C.Marino }

     \vskip 0.6cm
     
    \centerline{\it Instituto de F\'\i sica }
    \centerline{\it Universidade Federal do Rio de Janeiro }
    \centerline{\it Cx.P. 68528}
    \centerline{\it Rio de Janeiro RJ 21945-970}
    \centerline{\it Brazil}

\vskip 1.0cm

\begin{abstract} 
 
We consider the Maxwell-Higgs system in the broken phase, described in terms
of a Kalb-Ramond field interacting with the electromagnetic
field through a topological coupling. We then study the creation operators of 
states which respectively carry a point charge and a closed magnetic string
in the electromagnetic language or a point topological charge and a closed
Kalb-Ramond charged string in the Kalb-Ramond dual language.
Their commutation relation is evaluated, implying they
satisfy a dual algebra and their composite possesses generalized statistics.
In the local limit where the radius of the string vanishes, only Fermi or
Bose statistics are allowed. This provides an explicit operator realization
for statistical transmutation in 3+1D.
			      
\end{abstract}

\vskip 3cm
$^*$ Work supported in part by CNPq-Brazilian National Research Council.
     E-Mail address: marino@if.ufrj.br

\end{titlepage}

\hoffset= -10mm

\leftmargin 23mm

\topmargin -8mm
\hsize 153mm
 
\baselineskip 7mm
\setcounter{page}{2}

\section{Introduction}

The mechanism of statistical transmutation occurring 
in 2+1D for the composite state of a
point charge and a magnetic vortex \cite{fw} has been
generalized for 3+1D a few years ago. It was shown that the composite of 
a closed Nambu charged string and a 
point Kalb-Ramond vortex presented generalized
statistics \cite{gs} in four-dimensional spacetime. 
Also the Chern-Simons mechanism of statistical 
transmutation \cite{pol} was shown to hold for stringlike objects in 3+1D
\cite{fgt}. More recently, in the framework of the Abelian Higgs Model (AHM),
it was demonstrated 
that a closed magnetic (Nielsen-Olesen) string in the presence of
a point charge also displayed the phenomenon of Fermi-Bose transmutation 
\cite{fg}. In this work we provide an operator realization for
this fact. We consider the AHM in the broken phase, which
can be described in terms of an antisymmetric tensor (Kalb-Ramond) gauge
field interacting with the electromagnetic field by means of a 
topological coupling. In this language the topological charge of the 
Kalb-Ramond field becomes the electric charge and the Kalb-Ramond charge
becomes the magnetic flux. A creation operator, $\sigma$,       
for the magnetic string states of the AHM was recently introduced
in \cite{ms} and a creation operator, $\mu$,
for the topologically charged states of the Kalb-Ramond field was constructed
some time before in \cite{qedt}. Here, we obtain the operator dual to 
$\mu$, in the Kalb-Ramond framework and show that it is nothing but
the magnetic string operator $\sigma$, introduced in \cite{ms}  
rewritten in terms of the antisymmetric tensor
gauge field. We evaluate their commutation relation and thereby  explicitly 
show that the composite operator $\psi 
=\mu\sigma$ has generalized statistics. 
In the local limit where the radius
of the closed string vanishes, we show that only Fermi or Bose statistics
are allowed.
In Kalb-Ramond language these composite states carry both a point topological
charge (point Kalb-Ramond vortex) and a closed Kalb-Ramond charged string.
In electromagnetic language, they carry both a point electric charge and
a closed magnetic string. Through
the composite operator $\psi$, therefore, we obtain an explicit
realization for both the mechanisms discovered in \cite{gs} and \cite{fg}.

\section{The Maxwell-Higgs System and the Antisymmetric Tensor Gauge Field}

Let us consider the Abelian Higgs Model, given by the lagrangian
\be
\cl_{AHM} = -\fr{1}{4} F_{\mu\nu} F^{\mu\nu} + |D_\mu \phi|^2 + 
m^2 |\phi|^2 - \fr{\lambda}{4} |\phi|^4
\label{ahm}
\ee
Using the polar representation for the Higgs field $\phi = \fr{\rho}
{\sqrt{2}}e^{i\theta}$ and integrating over $\theta$ in the
approximation where the field $\rho$ is made equal to a constant $\rho_0$ 
(large $\lambda$), we
obtain the following lagrangian describing the effective dynamics of the
electromagnetic field in the AHM \cite{nv,ms},
\be
\cl [A_\mu] = -\fr{1}{4} F_{\mu\nu} \lef [ 1 + \fr{M^2}{-\Box} \ri ] 
F^{\mu\nu}
\label{la}
\ee
where $M = e\rho_0$. We can express any of the two terms above in terms
of an antisymmetric tensor gauge field (Kalb-Ramond field), by means
of the following identity in euclidean space
$$
\int DB_{\mu\nu} \exp \lef \{ - \int d^4z \fr{1}{12} H_{\mu\nu\alpha}
\lef [ \fr{A}{-\Box} \ri ] H^{\mu\nu\alpha} + \fr{1}{2} 
\epsilon^{\mu\nu\alpha\beta} A_\mu \del_\nu B_{\alpha\beta} + \cl_{GF}
\ri \} =
$$
\be
\exp \lef \{ - \int d^4z \fr{1}{4} F_{\mu\nu} 
\lef [  \fr{1}{A} \ri ] F^{\mu\nu} \ri \}
\label{id}
\ee
In the above expression, $H_{\mu\nu\alpha}$ is the field intensity tensor
corresponding to the Kalb-Ramond field $B_{\mu\nu}$, $A$ is an arbitrary
linear operator and $\cl_{GF}$ is the gauge fixing term. Making the choice
$A=\fr{- \Box}{M^2}$, we can reproduce the second term of (\ref{la}) and 
obtain the following equivalent lagrangian 
\be
\cl [A_\mu, B_{\mu\nu}] = - \fr{1}{4} F_{\mu\nu} F^{\mu\nu} -
\fr{1}{12} H_{\mu\nu\alpha} H^{\mu\nu\alpha} - 
\fr{1}{4} \epsilon^{\mu\nu\alpha\beta} B_{\mu \nu} F_{\alpha\beta}
\label{lab}
\ee
where we have rescaled the antisymmetric field as $B_{\mu\nu} \rightarrow
M \ B_{\mu\nu}$. This lagrangian describes the properties of the Abelian
Higgs Model in the constant $\rho$ approximation \cite{fg,lab}.
The operator field equations corresponding to it are
\be
\del_\alpha H^{\alpha\mu\nu} = \fr{M}{2} \epsilon^{\mu\nu\alpha\beta} 
F_{\alpha\beta}
\label{e1}
\ee
and
\be
\del_\nu F^{\nu\mu} = \fr{M}{2} \epsilon^{\mu\nu\alpha\beta} \del_\nu
B_{\alpha\beta}
\label{e2}
\ee
From (\ref{e2}) we obtain 
\be
\del_\alpha H^{\alpha\mu\nu} = \fr{\Box}{2M} \epsilon^{\mu\nu\alpha\beta} 
F_{\alpha\beta}
\label{e3}
\ee
Consistency between (\ref{e1}) and (\ref{e3}) implies that the 
electromagnetic field satisfies the Proca equation which describes the well
known screening of the broken phase of the theory.
We see from (\ref{e2}) that the topological
current of the antisymmetric field, namely,
$J^\mu = \fr{1}{2} \epsilon^{\mu\nu\alpha\beta} \del_\nu  B_{\alpha\beta}$
is proportional to the electric current:
$j^\mu_{el} = e\rho_0 J^\mu$. Also,
from (\ref{e1}), we conclude that the magnetic field is given by
$B^i = \fr{1}{M} \del_j \Pi^{ji}$, where  $\Pi^{ji} = - H^{0ji}$ is the
momentum canonically conjugated to the antisymmetric field $B^{ij}$.

\section{The String and the Charge Creation Operators}

Let us introduce now the operators which are going to create respectively
magnetic string states and charged states. The charge creation operator,
of course is the one which creates topological excitations, since 
the electric charge is identified with the topological charge. This has been
introduced in \cite{qedt}.
In terms of
the antisymmetric tensor field, it can written as
\be
\mu(\vec x,t) = \exp \lef \{ \fr{ib}{4\pi M} \int d^3  \vec \xi\  \del_i
\varphi_j ( \vec \xi -\vec x) \Pi^{ij} (\vec \xi, t) \ri \}
\label{m1}
\ee
or, in covariant form,
\be
\mu(\vec x,t) = \exp \lef \{ \fr{-ib}{4\pi M} \int d^3  \xi_\mu\  \del_\nu
\varphi_\alpha ( \vec \xi -\vec x) H^{\mu\nu\alpha} (\vec \xi, t) \ri \}
\label{m2}
\ee
In the above expressions $\varphi^\mu = (0,\vec\varphi)$ and $\vec \varphi =
\fr{1- \cos \theta}{r \sin \theta} \hat \varphi$,
$b$ is an arbitrary real parameter and a regulating procedure
is implicit \cite{qedt}. 
The equal-time
commutator of $\mu$ with the topological
carge (electric charge) $Q =\fr{M}{2} \epsilon^{ijk}\del_i B_{jk}$ was
evaluated in \cite{qedt}, yielding the result
\be
[Q,\mu] = b \mu
\label{qm}
\ee
This shows that the operator $\mu$ carries $b$ units of electric charge.

Let us introduce now the (closed) magnetic string operator. This is given
by
\be
\sigma(C,t) = \exp \lef \{ \fr{iaM}{2} \int_{S(C)} d^2 \xi_{\mu\nu}
\fr{\del_\alpha H^{\alpha\mu\nu}}{-\Box} \ri \}
\label{s1}
\ee
or
\be
\sigma(C,t) = \exp \lef \{ \fr{-iaM}{2} \int_{S(C)} d^2 \xi_{ij}
B^{ij} + gauge\ terms \ri \}
\label{s2}
\ee
In the above expressions, $a$ is an arbitrary real 
number and $S(C)$ is a space-like
surface bounded by the closed string at $C$.
$d^2 \xi_{ij}$ is the surface element of $S(C)$, the directions $i,j$ being
along the surface. The gauge terms in (\ref{s2}) guarantee the gauge
invariance of $\sigma$ which is explicit in (\ref{s1}).
Later on, it will become clear
that both the correlation functions and commutation rules of $\sigma$
are independent of the surface $S$: they just depend on $C$. 
The generalization for an open string is straightforward.
The operator
$\sigma(C)$ creates a magnetic string along the closed curve C. In order to
prove this, let us consider the magnetic flux operator along a surface
$R$. This is given by
\be
\Phi_R = \int_R d^2 \xi^i B^i = \int_R d^2 \xi^i  \fr{1}{M} \del_j \Pi^{ji}
\label{fl}
\ee
where we used the expression of the magnetic field $B^i$ in terms of the 
antisymmetric tensor momentum. Writing $\sigma \equiv e^\alpha$, we obtain
$[ \Phi_R, \sigma ] = \sigma [ \Phi_R, \alpha]$ since the latter commutator
is a c-number. Indeed, using (\ref{s2}), (\ref{fl}) and the canonical
commutation rules of the antisymmetric field, we immediately get
\be
[\Phi_R, \alpha] = a \int_R d^2 \eta^j \oint_C d \xi^j \delta ^3 (\xi -\eta)
\label{fia}
\ee
The above integrals give $\pm 1$ whenever 
the curve C pierces the surface $R$ in
the positive or negative sense, respectively. Otherwise they vanish. 
Hence, we if we choose the magnetic flux surface and the string in such a 
way that it pierces the surface positively, we get
\be
[ \Phi_R, \sigma ] = a\  \sigma
\label{fis}
\ee
This shows that the $\sigma$ operator carries $a$ units of magnetic flux 
along the curve $C$. Actually, substituting (\ref{e3}) in (\ref{s1})
we obtain precisely the magnetic string operator introduced in \cite{ms}.

\section{Dual Algebra and Statistical Transmutation}

Let us determine now the commutation rule between the operators $\sigma
\equiv e^\alpha$ and $\mu \equiv e^\beta$. Since $[\beta, \alpha]$ is a
c-number, we have
\be
\mu(\vec x, t) \sigma(C_y, t) \equiv \sigma(C_y, t) \mu(\vec x, t) 
e^{[\beta, \alpha]}
\label{ms}
\ee
Using (\ref{m1}), (\ref{s2}) and the canonical commutation rules of the 
antisymmetric tensor field, we get
\be
[\beta(\vec x,t), \alpha (C_y, t) ] = i \fr{ab}{8 \pi} \int_{S(C_y)}
d^2 \xi^i \epsilon^{ijk} ( \del_j \varphi_k - \del_k \varphi_j )
(\vec \xi -\vec x)
\label{ba}
\ee
where $d^2 \xi^i = \fr{1}{2} \epsilon^{ijk}  d^2 \xi^{jk}$ is the surface 
element of $S(C_y)$. The identity 
\be
(\del_j \varphi_k - \del_k \varphi_j )(\vec \xi -\vec x) \equiv
\epsilon^{ijk} \del_i \lef [ \fr{1}{| \vec \xi -\vec x |} \ri ]
\label{id}
\ee
allows us to obtain the result
\be
[\beta(\vec x,t), \alpha (C_y, t) ] = i \fr{ab}{4 \pi} \int_{S(C_y)}
d^2 \xi^i \fr{(x - \xi)^i}{| \vec x - \vec \xi |^3} =
i \fr{ab}{4 \pi} \Omega (\vec x; C_y)
\label{ba1}
\ee
where $\Omega (\vec x; C_y)$ is the solid angle comprised between $\vec x$
and the curve $C_y$. As a consequence we get
\be
\mu(\vec x, t) \sigma(C_y, t) = \exp \lef 
\{i \fr{ab}{4 \pi} \Omega (\vec x; C_y) \ri \}
\sigma(C_y, t) \mu(\vec x, t) 
\label{ms1}
\ee
This is an algebra analogous the order-disorder algebra 
\cite{kc,th,odd} and characterizes the charge and magnetic string
operators $\mu$ and $\sigma$ as dual to each other. Observe that (\ref{ms1})
is surface independent. Also the correlation functions of $\sigma$, in the
electromagnetic language, were shown to be surface independent \cite{ms}.

We can now construct the creation operator for the composite state carrying
charge and magnetic flux along a closed curve $C$. 
We can choose the curve $C_x$ to be a 
circle of radius $R$ centered at $\vec x$ 
and place the charge in the center, namely, 
\be
\psi(x;C_x;t) = \lim_{\vec x \rightarrow \vec y}
\mu(\vec x, t) \sigma(C_y, t)
\label{psi}
\ee
Using the fact
that $\Omega (\vec x; C_y) - \Omega (\vec y; C_x) = 4 \pi 
\epsilon \lef (\Omega (\vec x; C_y) \ri ) $, where $\epsilon(x)$ is the 
sign function, we obtain from (\ref{ms1})
\be
\psi(x;C_x;t) \psi(y;C_y;t) = e^{i\ ab \epsilon 
\lef (\Omega (\vec x; C_y) \ri )}
\psi(y;C_y;t) \psi(x;C_x;t)
\label{pp}
\ee
This relation shows that the composite charge-magnetic string state 
possesses statistics $s=\fr{ab}{2\pi}$ thereby providing an 
explicit operator realization for the statistical transmutation of 
strings in the presence of point sources, discovered in \cite{fgt,fg}.
In the local limit when we make the radius of the circle to vanish, 
the exponential factor becomes a constant and the only consistent 
possibilities for $ab$ are either $ab=\pi$ or $ab=2\pi$ (and their
corresponding multiples), meaning that $\psi$ can only be a fermion
or a boson in the local case. This agrees with the well known fact 
that only fermionic or bosonic local fields can exist in 3+1D.

\section{Remarks on Bosonization}

The operator construction presented here shows that it is possible 
to describe fermionic states in terms of bosonic gauge fields in 3+1D.
This is in agreement with the bosonization of the action and of the 
current in four dimensional spacetime
\cite{bos}. Of course, the crucial problem
of bosonization is to find the correct operators which would describe
the correlation functions of a specific fermionic thery in the framework
of the corresponding bosonic one. Bosonized lagrangians corresponding
to fermionic theories have been obtained \cite{bos}. The operators studied
here may be a step towards the obtainment of a full bosonization in 3+1D.
It is interesting to note that if this kind of construction would prove
to be correct, say, for the description of electrons, then these would 
come up as closed string states bound to a charge. The relation between
charge and magnetic flux responsible for the fermionic statistics would
at the same time explain the quantization of charge.

\leftline{\Large\bf Acknowledgements} \bigskip

I am grateful to R.Banerjee for calling my attention to reference \cite{fg}.

\end{document}